\documentstyle[aps,preprint]{revtex}
\def\be{\begin{equation}}
\def\ee{\end{equation}}
\def\bea{\begin{eqnarray}}
\def\eea{\end{eqnarray}}
\begin{document}
\begin{center}
{\LARGE \bf Black holes have no short hair}
\vskip1cm
{\bf Dar\'\i o N\'u\~nez, Hernando Quevedo, and Daniel Sudarsky}\\
\vskip5mm
Instituto de Ciencias Nucleares\\
Universidad Nacional Aut\'onoma de M\'exico \\
A. P. 70--543, M\'exico, D. F. 04510,  M\'exico. \\
\end{center}

\begin{abstract} 
{We show that in all theories in which black hole hair
has been discovered, the region with non-trivial
structure of the non-linear matter fields must extend
beyond $3/2$ the horizon radius, independently
of all other parameters present in the theory. We argue
that this is a universal lower bound that applies in
every theory where hair is present. This {\it no
short hair conjecture} is then put forward as a more modest alternative 
to the original {\it no hair conjecture}, the validity of which now seems
doubtful.} 
\end{abstract}

\vfill\eject

The canonical nature of stationary black holes, that is, the fact that they 
are completely specified by  the conserved charges that can be measured at 
asymptotic infinity, (like mass, angular momentum and electric charge ), was 
proven explicitly in Einstein vacuum theory, Einstein--Maxwell theory 
\cite{Wa}, and also for various types of theories involving scalar and 
vector fields \cite{Bek1}. The belief that this canonical nature had a 
universal validity, came to be known as the {\it no hair conjecture} 
(NHC) \cite{Whe}. This belief was based on the rigorous results mentioned 
above, and on the physical argument that suggested that all  matter fields 
present in a black hole space time would eventually be either radiated to 
infinity, or ``sucked'' into the black hole, except when those fields were
associated with conserved charges defined at asymptotic infinity.

The discovery of the ``color black holes" ({\it i. e.} static black hole
solutions in Einstein-Yang--Mills (EYM)\cite{eym} theory, that require for their
complete specification, not only the value of the mass, but also an additional
integer that is not, however, associated with any conserved charge), came as a
surprise, and it certainly forced us to reassess the status of the NHC.

The fact that these  solutions were afterwards proven to be 
unstable, seemed to allow a
resurrection of a more restrictive form of the NHC, which was, then, supposed to
apply only to stable black holes. The validity of this new version of the NHC
has become highly doubtful, to say the least, since the discovery of static
black hole solutions that are linear--perturbation--stable in theories like
Einstein-Skyrme (ES) \cite{es}, and apparently also 
Einstein--non Abelian--Procca (ENAP) \cite{eymh}, as
discussed in \cite{cat}. Thus, it seems clear now that there is very little 
hope for the validity of such a form of the NHC \cite{Biz2}.

The lack of validity of the NHC, naturally gives rise to the question: What
happened to the physical arguments put forward to support it? It seems 
clear now that the non-linear character of the matter 
content of the examples discussed plays an essential role:
The interaction between the part of the field that would be radiated away 
and that which would be sucked in is responsible for the failure of the
argument and, thus, for the existence of black hole hair.

On the other hand, this suggests that the non-linear behavior of the matter
fields must be present both, in a region very close to the horizon  (a region
from which presumably the fields  would tend to be sucked in)
and in a region relatively distant from the horizon (a region from which
presumably the fields  would tend to be radiated away), with the
self interaction being responsible for binding together the fields in these 
two regions.

We find convenient to introduce the term ``Hairosphere" to refer to the loosely
defined region where the non--linear behavior of the fields is present, in
contrast to the asymptotic region where the behavior of the fields is dominated
by linear terms in their respective equations of motion. A slightly more
explicit characterization of this region will be given below, after the proof of
our main result. 

The purpose of this letter is to show a result that gives support to the
heuristic argument mentioned above, by showing the existence of a lower bound
for the size of the ``Hairosphere". We do this by proving a theorem that applies
to all theories in which black hole hair 
has been found, and which states that this lower bound is parameter and theory
independent, and has the universal value given by $3/2$ the horizon radius.
Furthermore, the result also applies to black holes in theories whose matter
content is any combination of the matter fields corresponding to those theories.
Also, it is important to note that the existence of this lower bound is
particularly interesting in the cases where the fields in the theory are massive
(as in Einstein--Yang--Mills--Higgs (EYMH), ES, ENAP and Einstein--Yang--Mills--Dilaton (EYMD) with an
additional potential term \cite{dil}), as one could
have naively expected that, by adjusting the mass parameters of the theory, one
would obtain a black hole where the fields are substantially different from
zero, only within a neighborhood of the horizon that could be made as small as
one desires. Our result shows explicitly that this {\bf is not} what happens.

The first clues about the existence of this bound were obtained in \cite{cla},
where a procedure was presented, that allowed the construction of a Liapunov
function in Einstein-Higgs (EH) theory, giving a proof of a no--hair--theorem. When
applied to EYM theory, 
the same procedure yields instead a lower bound for the region of non--linear
behavior of the YM field. The present analysis is motivated by the fact that the
same procedure seems to be applicable to all theories known to exhibit hair, and
also by a new formulation \cite{Bek2} of the above mentioned no hair theorem in
EH theory, that was suggestive of a way to treat all interesting cases in a
unified fashion.

At this point, it seems convenient to define precisely what we call 
``hair":  We will say that, in a given theory, there is black hole hair 
when the space time metric  and the configuration of the other fields of a 
stationary black hole solution are not completely specified by
the conserved charges defined at asymptotic infinity. Thus, in the language 
of Ref. \cite{Pres}, we do not consider secondary hair.

We will focus on asymptotically flat static spherically symmetric black hole
space--times and write the line element as
\be  
ds^2 = -e^{-2\,\delta}\mu\,dt^2 +\mu^{-1}\,dr^2 +r^2\,(d\theta^2+\sin^2\,\theta\,d\varphi^2)\ ,
\label{eq:lel}
\ee 
where $\delta$ and  $\mu=1-2m(r)/r$ are functions of $r$ only, and we assume
that there is a regular event horizon at $r_H$, so $m(r_H)=r_H/2$, and
$\delta(r_H)$ is finite. Asymptotic flatness requires, in particular, that
$\mu \to 1$ and $\delta \to 0$, at infinity.

The matter fields will also respect the symmetries of the space-time, as is the
case, in particular, for the specific ansatz employed in each of the cases where
black hole hair has been discovered.

The Einstein's equations, ${G^\mu}_\nu=8\,\pi\,{T^\mu}_\nu$, together with the equations of motion for the field under consideration, form a dependent set, as they are related by the conservation equation ${T^\mu}_{\nu;\mu}=0$. We will use only two of the

 Einstein's equations and this last equation . 

The conservation equation has only one non--trivial component:
\be
{T^\mu}_{r;\mu}=0.\label{eq:con}
\ee

Einstein's equations give
\bea
\mu^\prime&=&8\,\pi\,r\,{T^t}_t + {{1-\mu}\over r},\label{eq:mu}\\
\delta^\prime&=&{{4\,\pi\,r}\over{\mu}}\,({T^t}_t-{T^r}_r ),\label{eq:delta}
\eea
where prime stands for differentiation with respect to $r$.

Using Einstein's equations (\ref{eq:mu}, \ref{eq:delta}) in (\ref{eq:con})
it is straight forward to obtain
\be
e^\delta\,(e^{-\delta}\,r^4\,{T^r}_r)^\prime={r^3\over{2\,\mu}}\left[
(3\,\mu-1)\,({T^r}_r-{T^t}_t) + 2\,\mu\,T \right],\label{eq:teo}
\ee
where $T$ stands for the trace of the stress energy tensor.

We will be making the assumption that the matter fields satisfy the weak energy
condition (WEC), that in our context means that the energy density, $\rho \equiv
-{T^t}_t$, is positive semi--definite and that it bounds the pressures, in
particular, $|{T^r}_r|\leq -{T^t}_t$.

Now we are in a position to obtain the following:

{\bf Theorem}: Let equation (\ref{eq:lel}) represent the line element of an
asymptotically flat static spherically symmetric black hole space time,
satisfying Einstein's equations, with matter fields satisfying the WEC and such
that the trace of the energy momentum tensor is non-positive, and such that the
energy density $\rho$ goes to zero faster than $r^{-4}$, then the function
${\cal E}=e^{-\delta}\,r^4\,{T^r}_r$ is negative 
semi-definite at the horizon and is decreasing between $r_H$ and $r_0$ where
$r_0 > \frac{3}{2}\,r_H$ and from some $r>r_0$, the function ${\cal E}$ begins
to increase towards its asymptotic value, namely $0$.

{\bf Proof:} First we note that the proper radial distance is given by
$dx=\mu^{-\frac{1}{2}}\,dr$, so equation (\ref{eq:teo}) can be written as 
\be
\frac{d}{dx}(e^{-\delta}\,r^4\,{T^r}_r)=\frac{e^{-\delta}
\,r^3}{2}\left[ \mu^{\frac{1}{2}}\,(3\,({T^r}_r-{T^t}_t) + 2\,T)
-\mu^{-\frac{1}{2}}\,({T^r}_r-{T^t}_t)\right].
\label{eq:pr1}
\ee

Since the components ${T^r}_r, {T^t}_t, {T^\theta}_\theta$
must be regular at the horizon, ({\it i. e.} the scalar
$T_{\mu\,\nu}\,T^{\mu\,\nu}=({T^r}_r)^2 + ({T^t}_t)^2 + 2\,({T^\theta}_\theta)^2
$ is regular at the horizon \cite{Bek2}) and, since $x$ is a good coordinate at
the horizon, the left hand side of equation (\ref{eq:pr1}) must be finite in the
limit $r \to r_H$, and since $\mu(r_H)=0$, we find:
\be
{T^r}_r(r_H) ={T^t}_t(r_H) =-\rho(r_H) \leq 0,\label{eq:equ}
\ee
so ${\cal E}(r_H)\leq 0$. Next, inspecting equation (\ref{eq:teo}), we note that
the right hand side is negative definite, unless $(3\,\mu-1)>0$. This follows
from the WEC, which requires $({T^r}_r-{T^t}_t)>0$, and the assumption that
$T<0$. Thus ${\cal E}
$ is a decreasing function at least up to the point where $3\,\mu -1$ becomes
positive, and this occurs at $r_1=3\,m(r_1)$, therefore
$r_0>r_1$. Since $m(r)$ is an increasing function, (as follows from the WEC and
the fact that eq. (\ref{eq:mu}) can be written as $m^\prime=4\,\pi\,r^2\,\rho
>0$) we then have:
\be
r_0>3\,m(r_1)>3\,m(r_H)=\frac{3}{2}r_H.
\ee
Q. E. D.

We see that, under the conditions of the theorem, the asymptotic behavior 
of the fields can not start before $r$ is sufficiently large since this 
behavior is characterized by the fact that ${T^r}_r$ approaches zero, at least as $r^{-4}$. And, in particular, in the asymptotic regime ${\cal E}$ is not simultaneously negative and decreasing.

The physical significance of the behavior of the function ${\cal E}$ can be seen from the following facts:

>From the inequality ${\cal E}(\frac{3}{2}r_H)<{\cal E}(r_H)\leq 0$,  and from the negativity of the radial pressure, ${T^r}_r$, it follows that
\be
|{T^r}_r(\frac{3}{2}r_H)|>{\left(\frac{2}{3}\right)}^4\,|{T^r}_r(r_H)|\,e^{(\delta(\frac{3}{2}r_H)-\delta(r_H))}.\label{eq:inq1}
\ee
>From the identification ${T^r}_r(r_H)={T^t}_t(r_H)$  in Eq.~(\ref{eq:equ}), and using the formula \cite{visser}:
\be
e^{-\delta(r_H)}=\frac{2\,r_H\,\kappa}{1+8\,\pi\,r_H^2{T^t}_t},
\ee
where $\kappa$ is the surface gravity of the black hole, we obtain:
\be
|{T^r}_r(\frac{3}{2}r_H)|>{\left(\frac{2}{3}\right)}^4 
\frac{2\,r_H\,\kappa}{1+8\,\pi\,r_H^2\,{T^t}_t(r_H)} \,|{T^t}_t(r_H)|\, e^{\delta(\frac{3}{2}r_H)}. \label{eq:bou1}
\ee
Next, note that from Eq.~(\ref{eq:delta}) and the weak energy condition, it follows that $\delta(r)>0$ (because $\delta(\infty)=0$). Then, we conclude that:
\be
|{T^r}_r(\frac{3}{2}r_H)|>{\left(\frac{2}{3}\right)}^4
\frac{2\,r_H\,\kappa}{1+8\,\pi\,r_H^2\,{T^t}_t(r_H)}\,|{T^t}_t(r_H)|. 
\label{eq:bou}
\ee

We would like to stress here that the bound (\ref{eq:bou}) on the value of the
radial pressure at $r=\frac{3}{2}r_H$ is expressed completely in terms of physical
quantities evaluated at $r_H$.
Thus, we interpret the theorem as stating that, under the conditions of its
hypothesis, the matter fields start their asymptotic behavior at some
$r>\frac{3}{2}\,r_H$, and thus that the ``Hairosphere" must extend beyond this point.

It is however worth pointing out that, although Eq.~(\ref{eq:bou}) has a clear physical interpretation, it is probably not the best bound that can be put on the value of ${T^r}_r(\frac{3}{2}r_H)$, since there is no general lower bound on the surface gravi

ty. Nevertheless, it seems clear that if the matter fields have affected the space-time structure in such a way as to dramatically alter the value of $\kappa$ ({\it i. e.} a Schwarzschild black hole has a value of $\kappa={1\over{2\,r_H}}$), then the hair

 can not be regarded as ``tiny". In fact, a much better bound on ${T^r}_r(\frac{3}{2}r_H)$ can be obtained by considering instead Eq.~(\ref{eq:inq1}): 
Integrating  Eq.~(\ref{eq:delta}) we find
\be
\delta({3\over2}r_H)-\delta(r_H)=4\,\pi\,{\int^{{3\over2}r_H}_{r_H}}\,{{r\,({T^t}_t - {T^r}_r)}\over{\mu}}d\,r
\label{eq:int}
\ee
one can then substitute the particular form of ${T^t}_t - {T^r}_r$ corresponding to the specific theory one is considering. The corresponding expressions are listed in table 1 for all theories known to exhibit hair, and since in all cases
${T^t}_t - {T^r}_r$ is proportional to $\mu$, we find that Eq.~(\ref{eq:int}) leads to an explicit bound on $\delta(r_H)-\delta({3\over2}r_H)$ in terms of the maximal value of the fields in the interval $[r_H, {3\over2}r_H]$. This in turn, when substitute

d in Eq.~(\ref{eq:inq1}), leads to an explicit bound on ${T^r}_r(\frac{3}{2}r_H)$ in terms of those fields. 

Another, perhaps more dramatic example of the physical significance of the result, is provided by the fact that, as a particular case\cite{TJac}, our theorem rules out the possibility of a realistic static shell (with finite thickness), made out of matter

 satisfying the WEC and the $T \leq 0$ condition, laying completely in the interval $[r_H, r_0]$, {\it i. e.} it can not be completely contained within the ``Hairosphere".

Next, we examine the relevance of the theorem. First we note that, in all
theories where hair has been found (EYM, ES, EYMH, ENAP, and
Einstein--Yang--Mills--Dilaton with or without and additional potential term
\cite{dil}), the conditions of the theorem hold as shown in table 1, so our
results apply to all these theories. Furthermore, the additivity of the energy
momentum tensor ensures that in a theory involving a collection of any number of
these fields (with the same type of ansatz), the condition $T\leq 0$ will continue to hold. The requirement that $r^4\,{T^r}_r$ goes to zero at
infinity, does not follow from asymptotic flatness, as can be seen from the fact
that it is violated for example by the Reissner-Nordstr\"om solution in
Einstein--Maxwell theory, but this is not a case where hair is present as there is an
additional conserved charge that is needed to complete the specification of the
solution.  In fact, the charges defined at asymptotic infinity are associated
with the $r^{-2}$ behavior of the fields and in general, the energy momentum
tensor is at least quadratic in these fields. Thus, this last requirement seems
to be the natural way to impose the condition that there are no extra charges
associated with the fields.

It appears then, that a suitable definition of the ``Hairosphere" could be to
take it as the complement of the region where $r^4\,{T^r}_r$ monotonically
approaches zero.

This set of results which have been obtained under the
assumption of spherical symmetry, in part because all
the cases in which hair has been discovered also involved this
simplifying assumption, should, we believe, generalize to the
stationary black hole cases where we expect that the 
"Hairosphere" should also be characterized by the length
$r_{Hair} = 3/2 \sqrt{A \over{4 \pi}}$ where $A$ is the 
horizon area.

In view of the evidence shown here, and based on the physical arguments
described at the beginning of this letter,
that suggest the existence of such a universal lower bound,
we are lead to conjecture that for all stationary black holes in theories
in which the matter content satisfies the WEC as well as the $T \leq 0$ condition, the ``Hairosphere'', if it exists,
must extend beyond the above mentioned distance. In short:
{\it If a black hole has hair, then it can not be
shorter that $3/2$ the horizon radius}.
 
It is interesting to note that if we take the natural expectation that in
theories with massive fields, stationary configurations will correspond to
fields that decrease rapidly within a Compton length of the horizon, {\it i. e.}
that the "Hairosphere" lies within $r_{hair}<\frac{1}{mass}$, and combine it
with the result presented above, we obtain an upper bound for the size of hairy
black holes, namely 
\be
r_H<\frac{2}{3}\,r_{hair}<\frac{2}{3}\,\frac{1}{mass}.
\ee

Thus, big black holes will have no hair. Actually, the numerical investigations
\cite{cat} of all massive theories known to present hair show
evidence for such an upper bound on the size of hairy black holes. This gives
support to the view that the present {\it no short hair} results can be taken as
an alternative to the no hair conjecture.

The validity of this conjecture as well as the general rigorous definition of
the "Hairosphere" should be matters of further research.

One of us, DS, wants to acknowledge helpful discussions with R. M. Wald.

\vfill
\eject

{\offinterlineskip
 \halign{ \strut
          \vrule#&
          \ \ \bf# \quad &
          \vrule#& \quad \hfil $#$ \hfil &
          \vrule#& \quad \hfil $#$ \hfil &
          \vrule#& \quad \hfil $#$ \hfil &
          \vrule#\cr
\noalign{\hrule} & && && && &\cr 
& && T && T^t_t - T^r_r &&  T^{t}_{t}(r \to \infty)  \sim 
&\cr & && && && &\cr \noalign{\hrule} & && && && &\cr
& Skyrme 
  && -{1 \over 16\pi}(f^2\,\mu\, {F^\prime}^2 
      + { \sin^4 F \over e^2\, r^4 })  
  && -{\mu \over 16\pi} (f^2 + { 2\sin^2F \over e^2\, r^2 }) 
      {F^\prime}^2 
  && {1 \over r^6} 
& \cr & && && && &\cr \noalign{\hrule} & && && && &\cr
&  YM 
   && 0  
   && -{\mu {w^\prime}^2 \over 2\pi f^2 r^2}
   && {1 \over r^6} 
& \cr & && && && &\cr \noalign{\hrule} & && && && &\cr
& YMD + V  
   && -{1 \over 4\pi}(\mu\,{\phi^\prime}^2 + 4V) 
   && -{\mu \over 4\pi} ({\phi^\prime}^2 
             + 2 {{w^\prime}^2 \over f^2\, r^2 } e^{2\gamma\phi}) 
   && {1 \over r^6} \,\,\, {\rm or}\,\,\, 
             {1 \over r^{10}} 
& \cr & && && && &\cr \noalign{\hrule} & && && && &\cr
& Y M H 
  && -{1 \over 4\pi}[\mu\,{\phi^\prime}^2 + 4V 
     + {{f^2\,\phi^2} \over {2\,r^2}}\,(1 + w)^2] 
  && -{\mu \over 4\pi} ({\phi^\prime}^2 
             + 2 {{w^\prime}^2 \over f^2\, r^2 } )
  && {e^{-2m\,r} \over r^2} 
& \cr & && && && &\cr \noalign{\hrule} & && && && &\cr
& NAProca 
  && -{ m^2 \over 8\pi\,r^2}\,(1 + w)^2 
  && -{\mu {w^\prime}^2 \over 2\pi f^2\, r^2}  
  && {e^{-2m\,r}\over r^2}  
&\cr & && && && &\cr \noalign{\hrule} } }

\vskip.2cm
{\bf Table 1.} The values of the trace, $T$, the difference 
$T^t_t - T^r_r$, and the asymptotic behavior of 
$T^t_t$, for the different theories known to have hair. 
The respective Lagrangian and ansatz are: For Skyrme 
${\cal L}_S=\sqrt{-g}\,(-{f^2\over 4}\,
tr( \nabla_\mu U \nabla^\mu U^{-1}) +
{1 \over {32\,e^2}}\,
tr[ (\nabla_\mu U)\,U^{-1}, (\nabla_\nu U)\,U^{-1}]^2)$, 
where $\nabla_\mu$ is the covariant derivative, $U$ is the 
$SU(2)$ chiral field, and $f, e$ are the coupling constants. 
For $U$ we use the hedgehog ansatz 
$ U(r) = e^{i\,({\bf{\sigma \cdot r}})\, F(r)} 
$ where ${\bf \sigma}$ are the Pauli matrices and 
${\bf r}$ is a unit radial vector. For Yang--Mills 
${\cal L}_{YM}=-{\sqrt{-g}\over 16\pi f^2} {F_{\mu\,\nu}}^a\,
{F^{\mu\,\nu}}_a$, where $f$ is the gauge coupling constant, 
we use the static spherically symmetric ansatz 
for the potential 
$A=\sigma_a\,{A_\mu}^a\,dx^\mu=\sigma_1 \,w\,d\,\theta 
+ (\sigma_3\,\cot\theta + \sigma_2\,w)\sin\theta\,d\phi$, 
where $w$ is a function of $r$ only. For Yang-Mills--Dilaton 
with potential 
${\cal L}_{YMD+V}={\cal L}_{YM}\,e^{2\,\gamma\,\phi}+
{{\sqrt{-g}}\over{4\,\pi}}
({1\over2}\nabla_\mu \phi \nabla^\mu \phi - V(\phi))$, 
where $\gamma$ is the
dimensionless dilatonic coupling constant, the ansatz for 
the gauge field configuration is
the same as that given in YM case, and $\phi=\phi(r)$. For
Yang--Mills--Higgs 
${\cal L}_{YMH}={\cal L}_{YM} -
{\sqrt{-g}\over 4\pi} \,((D_\mu \Phi)^\dagger\,(D^\mu \Phi) 
+ V(\Phi))$, where $D_\mu $ is the usual gauge--covariant 
derivative, $\Phi$ is a complex doublet Higgs field; 
the ansatz for the Yang--Mills field is the same as before, 
and for the Higgs field we have 
$\Phi={1\over\sqrt{2}}\left(\matrix{0 \cr \phi (r) \cr}\right)$, 
as usual $m=f<\Phi>.$ 
Finally, for Proca 
${\cal L}_{NAP}={\cal L}_{YM} - 
{{\sqrt{-g}\,m^2}\over{32\,\pi}}\,{A_a}^\mu\,{A_\mu}^a$, 
where $m$ is the mass parameter and the ansatz is defined as 
in the YM case.

\end{document}